\documentclass[prd,showpacs,twocolumn,aps,preprintnumbers,letterpaper,floatfix]{revtex4}
\usepackage{amsmath}
\usepackage{tipa}
\usepackage{amssymb}
\usepackage{graphicx,subfigure,amsbsy}
\usepackage{bm}
\usepackage{enumerate}
\usepackage{color}
\usepackage[export]{adjustbox}

\newcommand{\dd}{\mathrm{d}}
\newcommand{\be}{\begin{equation}}
\newcommand{\ee}{\end{equation}}
\newcommand{\bea}{\begin{eqnarray}}
\newcommand{\eea}{\end{eqnarray}}
\newcommand{\ba}{\begin{eqnarray}}
\newcommand{\ea}{\end{eqnarray}}

\newcommand{\GeV}{\,\mathrm{GeV}}
\newcommand{\MeV}{\,\mathrm{MeV}}
\newcommand{\fm}{\mathrm{fm}}
\newcommand{\mb}{\mathrm{mb}}

\begin{document}

\title{Collective interaction of QCD strings and \\ early stages of high multiplicity pA collisions}
\author{Tigran Kalaydzhyan}
\affiliation{Department of Physics and Astronomy, Stony Brook University,\\ Stony Brook, New York 11794-3800, USA}

\author{Edward Shuryak}
\affiliation{Department of Physics and Astronomy, Stony Brook University,\\ Stony Brook, New York 11794-3800, USA}

\date{\today}

\begin{abstract}
We study the early stages of ``central" $pA$ and peripheral $AA$ collisions. Several observables indicate that
at a sufficiently large number of participant nucleons  the system undergoes a transition
into a new ``explosive" regime. By defining a string-string interaction through the $\sigma$ meson exchange and performing
molecular dynamics simulation, we argue that one should expect a
strong collective implosion of the multi-string ``spaghetti" state, creating significant compression of the
system in the transverse plane. Another consequence
is the collectivization of the ``sigma clouds" of all strings into a
chirally symmetric fireball.
We find that these effects happen provided the number of strings $N_s>30$ or so,
as only such a number can compensate a small sigma-string coupling. These findings
should help us to understand the subsequent explosive behavior, observed for the particle multiplicities
roughly corresponding to this number of strings.
 \end{abstract}


\maketitle
\section{Introduction}

\subsection{The evolving views on the high energy collisions}

Before we get into our discussion of high-multiplicity $pA$ collisions, let us start by briefly
reviewing the current views on the two extremes: $AA$ and minimum-bias $pp$ collisions.

``Not-too-peripheral" $AA$ we define as those which have the number of participant nucleons $N_p > 40$,
and a corresponding multiplicity of the order of a few hundred. (Peripheral $AA$, complementary to this definition,
we discuss in Sec.~\ref{sec_peripheral_AA}.) Central $AA$ collisions produce
many thousands of secondaries: the corresponding  fireball has
an energy/entropy density well within the QGP domain, and these were naturally
in the ``mainstream" of the RHIC and LHC heavy-ion programs.
Needless to say, the theory guidance and those experiments resulted in
widely known conclusions about the strongly coupled dynamics of QGP. In particular, its
 collective flows were found to follow the hydrodynamical predictions with a remarkable accuracy.

(Hydrodynamical modeling  typically starts at the proper time $\tau_i\sim 1/2 \, \fm$, and
the equation of state used is that of the fully equilibrated
matter known from lattice simulations.
The description of matter at earlier stages and the exact
mechanism/degree of actual thermal equilibration are still a developing and hotly debated subject, which we do not
 address in this work.)

 AdS/CFT correspondence provides a dual description of strongly interacting  systems. In this vocabulary,
 the thermal fireball of deconfined matter is dual to a five-dimensional (5D) black hole, and its hydrodynamical
expansion corresponds to the motion of this black hole off the space boundary (where the gauge theory is located).
The attractively interacting and collapsing system of QCD strings we discuss should be viewed as a QCD
analog to AdS/CFT black hole formation.

  \begin{figure}[t]
  \begin{center}
  \includegraphics[width=6cm]{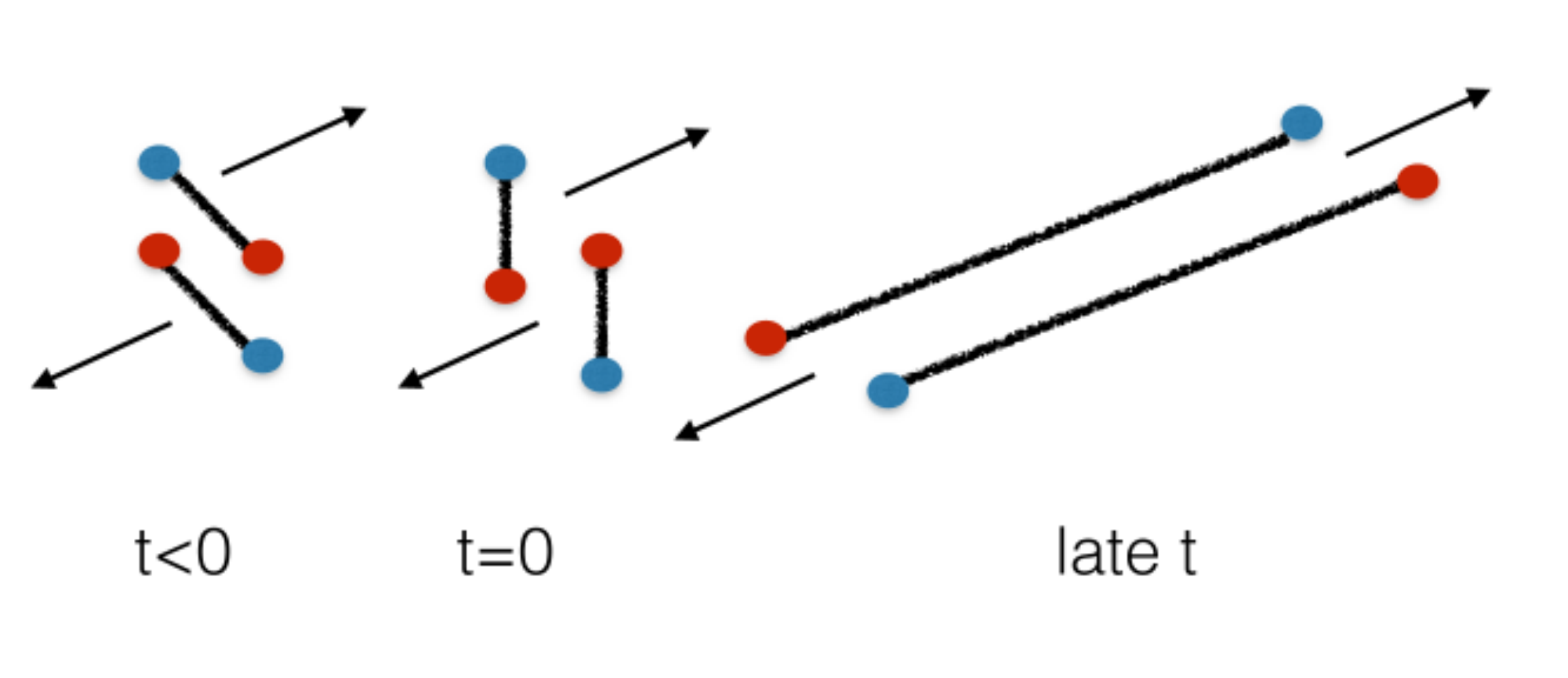}
  \includegraphics[width=6cm]{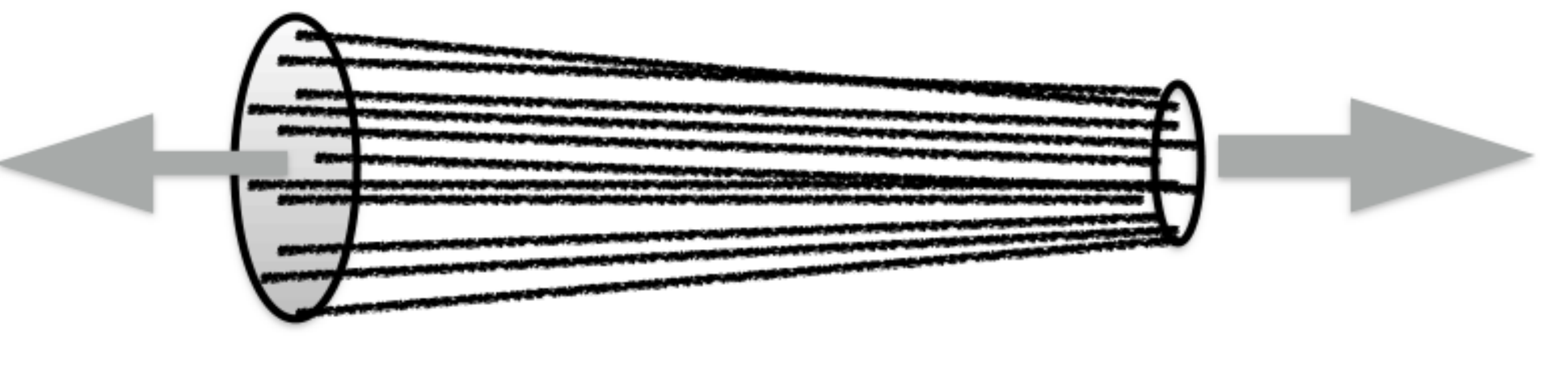}
  \caption{ (Color online) Top: Basic mechanism of a two-string production, resulting from color
  reconnection. Bottom: Sketch of the simplest multi-string state, produced in $pA$ collisions
  or very peripheral $AA$ collisions, known as ``spaghetti". }
  \label{fig_spaghetti}
  \end{center}
\end{figure}

   The opposite extreme is represented by typical (minimum bias) $pp$ collisions. Its Pomeron description
   at a large impact parameter $b=1-2\, \fm$ is naturally given in terms of a double-string production (see Fig.~\ref{fig_spaghetti}, top). Color reconnection (described perturbatively or semiclassically, by a ``tube" string diagram) leads to a pair of longitudinally  stretched strings,  with subsequent
  breaking into several pieces -- hadronic clusters, which finally decay into a few final secondaries, as implemented in, e.g., the Lund model event generators, which do quite a good job in reproducing these phenomena.
      The density of a produced excitation is low, and it takes place in the confining QCD vacuum: thus the string description.
   The Pomeron profile, in particular, was historically the origin of the so-called $\alpha'(t=0)$ parameter, related to the string tension, which defines the ``string scale" both in QCD and in fundamental
    string theory.

(If collisions  are ``hard", with a momentum transfer $Q\gg 1 \, \GeV$,  they resolve nucleons and Pomerons to their partonic substructure.
  The perturbative description of the Pomeron is  well developed. At a very high density perturbative
  theory breaks down and may lead to the ``color glass condensate" described by classical Yang-Mills fields.
  We, however, do not discuss hard collisions
  in this work, focusing instead on a soft string description.)

    Both of these very different descriptions have reached sufficient maturity by now.
   The main question we discuss in this work is how they can be reconciled. In doing so, we address collisions which are
    intermediate between these two extremes. More specifically, one may ask where exactly the transition takes place and how sharp it is.
    It was argued \cite{Shuryak:2013sra} that these two regimes are in fact {\em separated by a third}
   distinct regime similarly to the finite-temperature QCD, in which there is the so-called ``mixed" phase between the confined (hadronic) and the deconfined (partonic) phases. As known from decades of theoretical and numerical (lattice) research,
   this is  described most naturally by the {\em near-critical strings}, namely, strings in the Hagedorn regime.
 A similar transition
 has been found  \cite{Shuryak:2013sra}  in the framework of the holographic Pomeron:
 here string fluctuations can be described by a thermal theory, with the effective temperature directly related to the impact parameter $b$. All three phases -- the stringy one at large $b$, near-Hagedorn one at intermediate $b$, and
 (saturated) partonic one at small $b$ -- can indeed be seen in the elastic scattering amplitude profile $F(b)$.
 However, this theory has so far been developed only for the tunneling (Euclidean) stage of the system:
 more work is needed to describe the subsequent evolution of the system and to relate
 these three regimes to the observations in the corresponding $inelastic$ collisions.

 In this paper, instead of challenging ``central" (high multiplicity) $pp$ collisions, we focus on simpler problems:
 (i) high-multiplicity $pA$ and (ii) peripheral $AA$ collisions.
In both cases  the cross sections used to be in the standard Pomeron regime (impact parameters
as large as possible).
 So the number of participating nucleons $N_p$ and thus produced
 strings  can be calculated from a known cross section by the standard Glauber-type calculation.

 Yet if the number of involved participant nucleons
 is chosen to be large enough, $N_p= \mathcal{O}(10)$, the emerging configuration resembles the ``spaghetti" state shown in Fig.~\ref{fig_spaghetti} (bottom).
  The system of strings, once produced by color exchanges as the target and projectile pass each other at $t\approx 0$, is then stretched longitudinally between
the beams remnants, with large rapidities $+Y$ and $-Y$ comparable to those of the initial beams. As the generic
rapidity interval is large, $\geq 10$, we focus on some intermediate $y$
between, $-Y \ll y \ll Y$ (not  close to the ends), where the collider detectors are.
So one can view the set of strings as approximately parallel, directed along the beam direction.

 We study conditions under which the strings  are {\em no longer independent} from each other, and one has to take into account their  interaction and the collective effects it may produce.
  As we describe in detail, these effects change from (i) the simple ``spaghetti stage", in which all strings are simply parallel and independent,
  to (ii) the situation in which strings start to move due to the influence of others, all the way to (iii)
   $implosion$ into a localized cluster (or few clusters).

 \subsection{Explosive high multiplicity pp/pA collisions}

 The change in dynamics should lead to some visible changes in the observables. And indeed,
 as discovered at the LHC, low- and high-multiplicity  $pp$/$pA$ collisions show quite  different
 behaviors. But before we discuss recent LHC experiments, let us briefly review the history of the subject.

   The  idea that a very small system produced in $pp$ collisions, which can only be about $R_\perp\sim 1 \, \fm$ in its transverse
   size, can and should be treated hydrodynamically was first expressed by Landau in his pioneering 1953 paper \cite{Landau:1953gs}.
The road from this original idea to its realization turned out to be about 60 years long.
 Radial flow effects
were searched for  in minimum-bias $pp$ collisions at CERN ISR more than 30 years ago by one of us \cite{Shuryak:1979ds}, with negative results. The ISR spectra of identified ($\pi,K,p...$) secondaries possessed the
so called ``$M_\perp$-scaling",  consistent with independent string fragmentation and putting a rather stringent restriction on the possible magnitude of the
collective flow. As we discuss below, minimally biased $pp$ collisions do indeed
produce a very dilute string system which provides no pressure gradient needed for
an explosion.

In the 1990's Bjorken suggested that one can look not at the typical  (minimal-bias) $pp$  collisions, but at a
specially triggered high-multiplicity subset.
 Some indications for the radial flow in $\bar{p} p$ collisions at FERMILAB
 were found in higher multiplicity  MINIMAX events \cite{Brooks:1999xy}. Yet
 the magnitude of it remained inconclusive, the flow signal was weak, and  we are unaware of any actual
 comparison between the hydro flow and the MINIMAX data.

    With the advent of the LHC era of an extremely high luminosity and short-time detector capabilities,  a hunt for
   strong fluctuations in the parton multiplicity became possible.
   Already during the very first run of the LHC in 2010, the CMS collaboration was able \cite{CMS:2012qk} to
   collect a sufficient sample of high-multiplicity $pp$ collisions occurring with a probability of $\sim 10^{-6}$. CMS found the ``ridge"
   correlation in the highest multiplicity bins,
an angular correlation in the azimuthal angle between two particles
  at $\Delta\phi<1$ which extends to a large rapidity range $|\Delta y| \geq 4$.

 More recently the same phenomenon was seen in $pPb$
collisions as well, now by the CMS \cite{CMS:2012qk}, ALICE \cite{Abelev:2012ola}
 and ATLAS \cite{ATLAS} collaborations, as well as by PHENIX \cite{Adare:2013piz} in $dAu$ collisions at RHIC.
Larger number of ``participant nucleons" and higher average multiplicity
 substantially weaken the cost of the trigger: the ``ridge" is seen at the trigger level of a few percentages higher multiplicity events.
It was shown  that the ridge always comes with an ``anti-ridge" on the other side
of the azimuthal circle and is well described  by the second and third harmonics of  the azimuthal angle, very similar to that
in peripheral $AA$ collisions.  These ``elliptic" and ``triangular" flows are characterized experimentally by the parameters
$v_n=\langle \cos(n\phi)\rangle,$ $n=2,3$.

The geometry of the initial state is different in $pPb$ collisions at the LHC and $dAu$ collisions at the RHIC:  deuterons create ``two-center explosions",
 with an
elliptic deformation $\epsilon_2$ about twice as large \cite{Bozek:2011if}. The  observed flow is also about twice as large: $v_2^{dAu}/v_2^{pPb}\approx  2$.

 The radial flow in $pA$ collisions was predicted \cite{Shuryak:2013ke} to be even larger than in AA. This was subsequently verified by the
 spectra of identified secondaries $\pi, K$, and $p, \Lambda$ by CMS and ALICE.

  Recently the PHENIX $dAu$ data set has been analyzed for the HBT radii \cite{Adare:2014vax}:
 their modulation with the elliptic flow $v_2$ direction, and radius-momentum correlations, further
 support the presence of strong radial and elliptic flows.

All those observations lead to the conclusions that Landau's vision is perhaps realized in $pA$ collisions, with $P_{pA}\sim 10^{-2}$ probability in the highest multiplicity bins, and perhaps even in $pp$, with the much lower probability  $P_{pp}\sim 10^{-6}$. The question is what exactly happens in those events,  why
they are different from the minimum-bias ones. This paper is an attempt to answer some of these questions.

Note that the very fact that such a small system with a radius of $\sim 1\, \fm $ can behave macroscopically is not, by itself,
so surprising.
 After all, the matter produced, known as
sQGP, is approximately conformal. Its entropy density and viscosity scale as
\begin{align}
s\sim \eta \sim T^3
\end{align}
so they do not have any scale of their own. So a  small ``conformal copy" of a large  ($AA$) fireball should explode
in exactly the same way.
The problem is that, taken naively, these systems are not a ``conformal copy": their energy/entropy
density is high, but not high enough. Hydro calculations using naive initial conditions underestimate
the radial and elliptic flows.

A few comments on the pressure/entropy are in order. A QCD string (or a flux tube) is an object which is balanced in the transverse plane,
with all fields exponentially decreasing from its center. A ``spaghetti state" with parallel strings is not expected to
produce transverse pressure: in fact we argue below that it, rather, has a tendency to implode. Therefore, in order to
justify the observed explosive behavior of high-multiplicity events one needs to understand how a ``spaghetti state"
may turn itself into equilibrated QGP. The entropy issue is also related to this.
 If the hydrodynamical description of the explosion works (and it does), it tells us  that basically the entropy
 is approximately conserved at this stage. So it must be there from the onset of the hydro stage.
  A ``spaghetti state", on the other hand, does not have a high entropy: thus some important changes must have occurred in between.

\section{Setting the stage}


  \begin{figure}[t]
  \begin{center}
  \includegraphics[width=6cm]{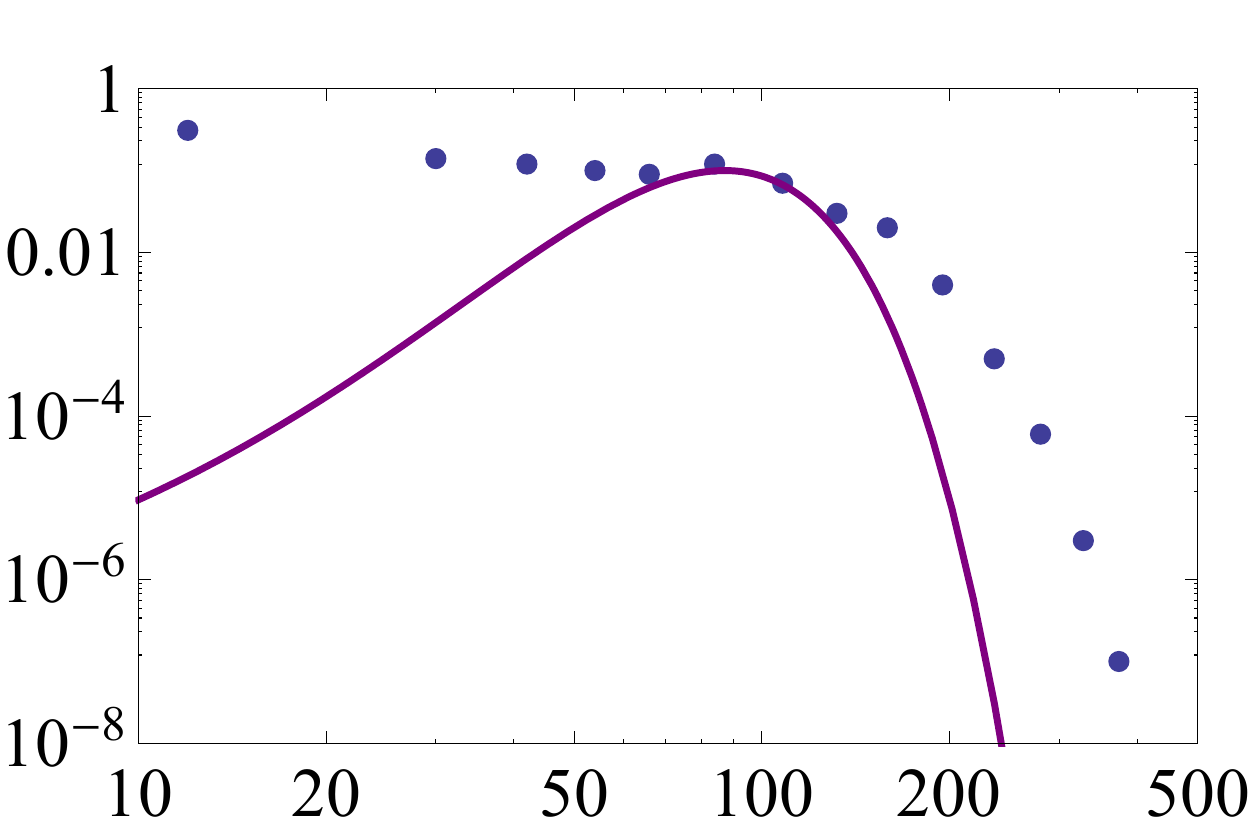}
  \caption{(Color online) Probability distribution over the number of charged tracks in the CMS detector acceptance $P(N_{tr})$ \cite{Chatrchyan:2013nka}.
  The (purple) line is the Poisson distribution with $\langle N_p \rangle=16$, arbitrarily normalized to
  touch the data points. }
  \label{fig_pA_mult}
  \end{center}
\end{figure}

\subsection{ A brief introduction to high multiplicity $pA$}
It is by now well known that the multiplicity distribution in $AA$ and $pA$ collisions is mostly
reproduced by the nuclear geometry and the so-called ``wounded nucleon" model.
It  assumes that the final particle multiplicity
(below characterized by the number of charged tracks $N_{tr}$  seen by the CMS detector, with $p_\perp>0.4\, \GeV$ and in certain rapidity window) is generated via
the distribution in the number of the so-called participant nucleons $N_p$.
Each participant nucleon is connected to the system by certain number of strings -- 2 as a minimum --
which fragment into secondaries independently.
Furthermore, for simplicity, we assume that this decay process has a narrow (or zero) dispersion, or
\begin{align}
{N_{tr} \over N_p}=const\,,
\end{align}

This simple one-parameter model is known to reproduce the multiplicity distribution in $AA$ collisions.
Glauber calculations for  $pA$ and $dA$ collisions allow us to calculate the distribution over $N_p$ (see, e.g.,
those reported by Bozek
\cite{Bozek:2011if}).
One, in fact, does not need a detailed simulation to understand the main numbers involved. A nucleon hitting a heavy nucleus has a  column of nuclear matter in front of it,
which varies in size and density. The $maximum$ obviously corresponds to the nuclear diameter $2 R_A$ with a density approaching the saturated density of nuclear matter
$n_0=0.16 \, \fm^{-3}$. A tube with cross section $\sigma$ contains an average number of nucleons
\begin{align}
\langle N_p \rangle=\sigma_{in} 2R_A n_0\approx 20\,,
\end{align}
where we have used the inelastic cross section at the LHC $\sigma_{in} \approx 100 \, \mb=10\, \fm^2$.
Accounting for a realistic density distribution in nuclei, one gets a slightly reduced number,
\begin{align}
\langle N_p \rangle= 16\,.
\end{align}
Furthermore, if one ignores any  correlations in nucleon positions (assumed  in all Glauber studies we are aware of),  the distribution in the number of participant nucleons $N_p$ should be described by a Poisson distribution with such an average.

In order to see if this simple model works for $pA$ collisions,  we plotted as points
the actual multiplicity distribution as measured by CMS (see Fig.~\ref{fig_pA_mult}).
(In the CMS terminology, ``tracks" are charged secondaries with $p_\perp>0.4\, \GeV$
in the rapidity coverage of their TPC tracker. The actual number of charged secondaries is
about twice that number, and taking unobserved neutral secondaries into account the true
multiplicity of secondaries produced is about a factor of 3 larger than the $N_{tr}$ used in this section.)

The curve in Fig.~\ref{fig_pA_mult} is
the Poisson distribution corresponding to the distribution over $N_p$ for a proton moving along the
$Pb$ nucleus diameter: it does not describe the data.  The left branch ($N_p  < \langle N_p \rangle$) is in fact explained
if a better account of the nuclear geometry -- columns with fewer nucleons --  is made (see, e.g., \cite{Bozek:2011if}).
(His Figs.~2 an 3 use the highest multiplicity cutoff at  the probability level of a few percent: specifically, $N_P>18$, 4 percent  centrality for $pPb$ and  $N_P>27$, 5 percent centrality for $dPb$.)

The right branch, with a larger multiplicity, $N_p  > 80$,  is different.
All fluctuations in the positions of the nucleons are already included in the curve:
 apparently, it is not enough to explain the tail.
Even if we relax the assumption of the zero-width decay of the strings and assign it a Poisson probability (still keeping the independence of the breaking processes), we will not describe the actual multiplicity pattern, which can be fitted by the negative binomial (NB) distribution (see, e.g., \cite{Kozlov:2014fqa}), because the negative binomial distribution itself is not a convolution of two Poisson distributions.
 We think that the wounded nucleon model
is inadequate, because for such multiplicities
there are
collective effects, described below.

The typical (inelastic) cross section of NN collisions at LHC energies is $\sigma_{in}\approx 10 \, \fm^2$. The
 impact parameter in a collision is thus
\begin{align}
\bar{b}\sim \sqrt{ {\sigma_{in} \over \pi}} \approx 1.5\, \fm\,, \label{lbar}
\end{align}
which is about 10 times the string radius  $r_s\approx 0.15\, \fm$.
The diluteness of our ``spaghetti" state - the fraction of the volume occupied by $N_s$ strings -- is then
\begin{align}
\langle \mathrm{diluteness} \rangle= N_s \left({r_s \over \bar{b}}\right)^2 \sim 10^{-2} N_s\,.  \label{diluteness}
\end{align}
For a ``minimally biased" (typical) collision producing just a few strings, it is indeed a rather dilute system. So, the independence of string fragmentation -- assumed by
the Lund model  -- seems to be reasonable. Even for $pA$ events corresponding to the heavy nucleus diameter, with $N_s \sim 30$, there seems to be enough space for all the strings, as the ``diluteness'' is only $0.3$.
And yet, as we show, there are good reasons to  revisit the
assumption of string independence.

 \subsection{The onset of radial, elliptic and triangular  flows}
  We do not review here any characteristics  of the radial, elliptic, and triangular  flows in most central $pA$
  and peripheral $AA$, as done elsewhere. Let us just say that for the same multiplicity
  (or, perhaps, for the same ``diluteness'' \cite{ropes_pheno}),
   they show striking similarities, which can be made even more precise using various scaling arguments, as in \cite{Lacey:2013eia} and \cite{Basar:2013hea}.
    For the purposes of this paper we only  need to localize the onset of all those phenomena.

{\bf Radial flow} modifies the shapes of particle spectra differently, depending on their mass $M$.
The data sets for  identified secondaries $\pi,K,p$, and $\Lambda$ show a
  specific dependence on the particle mass  of either (i) their mean transverse momentum $\langle p_\perp(M) \rangle$ or (ii) the
 $M_\perp$ distribution inverse slope $T'(M)$.
 The data (e.g., \cite{Chatrchyan:2013eya}) show no $M$ dependence for lower multiplicities,  but the effect appears for higher ones, $N_{tr}>80$.

 {\bf Elliptic flow} is, in these cases, measured also in two ways, by either the two-particle or the four-particle correlation parameters, $v_2\{2\}$ and $v_2\{4\}$. The latter one for $pA$ is
 especially sensitive to collectivity of the elliptic flow, and it has now been measured. It, however,
 rapidly drops below
 $N_{tr} \approx 80$ (see \cite{Chatrchyan:2013nka}). This is, perhaps, the best indicator for the onset of the explosive regime that
 we have so far. The $AA$ data for $N_{tr}<80$ are too uncertain to see any trends there.

 To summarize this subsection, there are at least three phenomena which seem to appear at about the  same multiplicity (in the CMS definition, $N_{tr}>80$), namely, (i)  the radial and (ii) elliptic flows, as well as (iii)
a multiplicity distribution  extending well beyond  that of the  ``wounded nucleon" model.

 \section{Collective string interactions}

 \subsection{Interaction in multi-string systems}

Systems of many QCD strings have been studied since the early 1990's in the context of Lund-like models of particle production.

 The first discussed effect was the possibility of having overlapping strings, or string fusion, in which the color fields of the strings add up and result in a stronger field \cite{ropes}. Decay of such objects, sometimes called ``color ropes", can enhance the production of strangeness and change $p_T$ spectra (see, e.g., \cite{ropes_pheno} and subsequent literature). The strings did not interact with each other in the sense we employ in this paper.

As the number of strings and their density increase, they can be merged into larger clusters.
The idea of {\em color percolation} was discussed in connection with deconfinement \cite{percollation1} and, also, strings \cite{Braun:1999hv}.
When the diluteness of strings reaches a certain critical
value, the cluster takes over the whole system, which was suggested to become ``deconfined" as a result.
For a recent relation to high-multiplicity $pp$ see also \cite{percollation2}.
(These ideas are, in a way, complimentary to our treatment below, which focuses, however, on chiral symmetry restoration
 rather than deconfinement.)

Studies of the interaction of $non$overlapping strings have been done on the lattice, in a different setting.
We do not go into this subject, because one can find a brief review of it
in our previous paper \cite{Kalaydzhyan:2014tfa}.

Unlike that paper, here we consider not complicated string shapes, ``string balls", but a set of parallel and straight (unexcited)
strings, called spaghetti. These strings have charges at the end, which move along the beam line, into positive and negative
directions, with high rapidities. The equation of motion will be derived for a small segment of the string in the middle, of length $\delta x$.
(The value of $\delta x$ is irrelevant, as it appears both in the mass and in the force, and cancels out.) The calculation is done in the nonrelativistic  approximation, assuming transverse string velocity $v\ll 1$ and ignoring retardation effects. The string-string
potential depends on $m_\sigma r$ and, therefore, has an uncertainty in the sigma mass (due to its large decay width).
This uncertainty is presumably larger than or comparable to the effect of retardation.

Following this, we assume the string interaction to be mediated by the lightest scalar $\sigma$. For one string the sigma ``cloud"  has a shape given by the quark condensate,
\begin{align} {\langle \bar q q(r_\perp) W \rangle \over \langle W \rangle \langle \bar q q \rangle} = 1-  C K_0( m_\sigma \tilde{r}_\perp)\,, \label{eqn_pot}
\end{align}
 where $K_0$ is the modified Bessel function and the ``regularized" transverse distance $\tilde{r}_\perp$ is
 \begin{align}
 \tilde{r}_\perp= \sqrt{ r_\perp^2+s_{string}^2}\,,  \label{eqn_tilde_r}
 \end{align}
 which smooths the two-dimensional (2D) Coulomb singularity $\sim \ln(r_\perp)$ at small $r_\perp$. The parameter $s_{string}$ is considered to be an ``intrinsic'' string width (in contrary to the effective string width, which is a result of quantum fluctuations).

   \begin{figure}[!h]
  \begin{center}
  \includegraphics[width=6cm]{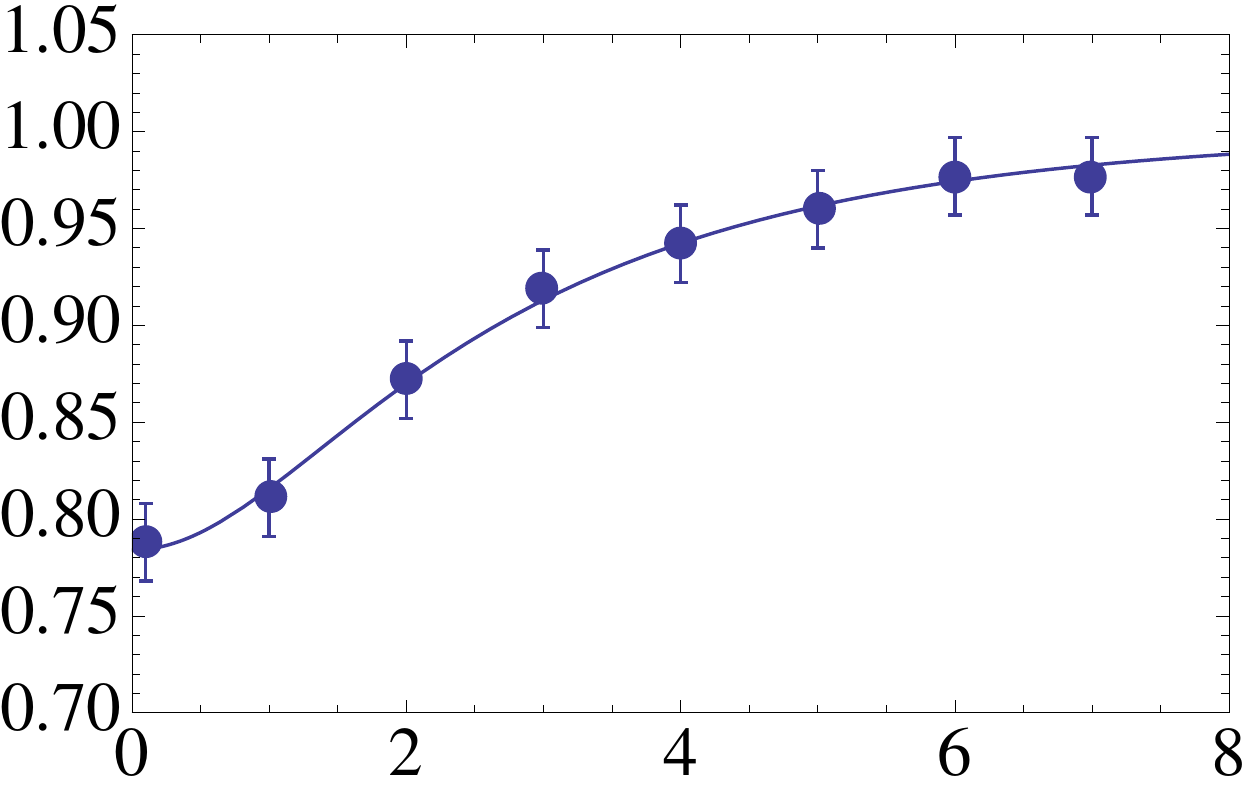}
  \caption{(Color online) Normalized chiral condensate as a function of the radial coordinate transverse to the QCD string. Points are from the lattice data \cite{Iritani:2013rla}. The curve is expression (\ref{eqn_pot}) with $C=0.26$ and
  $s_{string}=0.176\, \fm$.}
  \label{fig_lat}
  \end{center}
\end{figure}

 Lattice simulations, such as \cite{Iritani:2013rla}, have found vacuum modifications due to the presence of a QCD string.
 We argued \cite{Kalaydzhyan:2014tfa} that these data can be well described by a
 ``sigma cloud" (\ref{eqn_pot}). In Fig.~\ref{fig_lat} one can see our two-parameter fit to these data (the sigma mass here was taken to be  $m_\sigma=600\, \MeV$ as an input, and was not fitted/modified). According to our fit, the intrinsic width is $s_{string}\simeq 0.176\,\fm$. After we had done the calculations, other lattice data \cite{Cosmai} appeared, reporting the ``intrinsic'' width (penetration length of the dual superconductor picture in their case) to be equal to $\lambda=0.175\,\fm$. Taken that our calculation was not supposed to be so precise, we treat this as an amazing match.

Since the strings are almost parallel to each other, the problem is reduced to the set of
point particles on a plane with the 2D Yukawa interaction. From fit (\ref{eqn_pot}) one can see \cite{Kalaydzhyan:2014tfa}, that the main parameter of the string-string interaction (in string tension units) is numerically small,
 \begin{equation}
 g_N \sigma_T = \frac{\langle \sigma \rangle^2 C^2}{4 \sigma_T} \ll 1\,,
 \end{equation}
 typically in the range $10^{-1}- 10^{-2}$.
 So, it is correctly neglected in situations for which the Lund model was originally invented --
 when only $\mathcal{O}(1)$  strings are created.

 The interaction starts to play a role  when this smallness
 can be  compensated by a large number of strings. As shown in Fig.~\ref{fig_lat}, the magnitude of the quark condensate  $\sigma= |\langle \bar q q \rangle|$ at the string
 position is suppressed by about  $20\%$ of its vacuum value.
So, in a ``spaghetti" state one should think of the quark condensate suppression as being about 0.2 times the diluteness,
  (\ref{diluteness}), which is still $< 1$.

 On the other hand,  about 5 overlapping strings would be
 enough to eliminate  the condensate and completely restore the chiral symmetry.
 If $N_s > 30$ strings implode into an area several times smaller than $\sigma_{in}$, occur basically on top of each other and act coherently (which is the case as we will argue below), then  the chiral condensate will be eliminated inside a
 region of  1~$\fm$ in radius, or about 3~$\fm^2$ in area.
 This will create a small hot QGP fireball.

 Another possible approach to the flux tube interaction is developed
  within the Abelian Higgs (or dual superconductor) model of the QCD vacuum. In this picture the scalar field $\phi$ is produced by the condensed magnetic monopoles. The vector (electromagnetic) field acquires mass due to the nonzero vacuum expectation value of the scalar field $\langle \phi \rangle \neq 0$. The ratio of the masses $M_A/M_\phi$ defines the leading interaction channel between (and hence the sign of the interaction between) the Abrikosov flux tubes: if it is larger than 1, then one has domination of the attraction, i.e., a ``type-I" superconductor. We assume that this is the case. There are also recent lattice data \cite{Cosmai, Cea:2012qw} supporting our assumption in the $SU(3)$ case.

 \subsection{Other string phenomena}

Although we study neither string breaking nor the excitation and annihilation of strings, we should comment on
the parameters of these processes, since these will limit the time range of our description.

The longitudinally stretched strings in Fig.~\ref{fig_spaghetti} (top) eventually
break due to the so-called Schwinger-type mechanism of $\bar{q}q$ production
in the electric flux tube.
Without going into detail, we remind the reader that this
 implies an exponential suppression
of string breaking. Its typical proper time,
\begin{align}
\tau_{breaking} \sim 1 \, \fm/c\,,   \label{breaking}
\end{align}
may appear ``normal", but
is, in fact, a long time in string units.
 The clusters produced in the process of string breaking have a length of about $l_{breaking}\sim 2\,  \fm$,
 which is indeed large in comparison to the string radius of only about $0.15 \, \fm$.
A typical mass of the string segments is
$M_{breaking}\sim \sigma_T \cdot l_{breaking}\sim 2 \, \GeV$, so these are
not hadrons but ``stringy clusters", decaying into for or five pions later.

 After breaking, the string enters
the so-called yo-yo stage, in which the quark and antiquark at the string ends are
accelerated towards each other by the string tension. The effective transverse mass
of a quark is $M_\perp=\sqrt{m^2+p_\perp^2}\sim 1/2 \, \GeV$, while the string tension is
$\sigma_T\approx 1 \GeV/\fm$: thus, the effective quark acceleration $a_q\sim 2 \, \fm^{-1}$.
The quark rapidity changes as
\begin{align}
y=a_q \tau\,,
\end{align}
and this process happens from both ends of the string pieces, each about 2 units of rapidity
long. Therefore, the strings shrink and disappear within a time
$\tau_{yoyo}\sim 0.5 \, \fm/c$.
 Their sum,
 \begin{align}
 \tau_{breaking}+   \tau_{yoyo}\approx 1.5 \, \fm/c\,,
 \end{align}
  is naturally a limit of the time which is available for the collective string effects we study.

The presence of many strings of the same color and opposite color fluxes may induce
another dissipative phenomenon, namely, flux-antiflux annihilation.
It is also suppressed, not exponentially, but only by $N_c$.
If this happens, then the strings are reconnected as shown:
\be
\setlength{\unitlength}{.015in}
\begin{picture}(180,20)
\linethickness{0.15mm}
\put(00,00){\circle*{3}}
\put(00,00){\line(1,0){70}}
\put(00,20){\circle*{3}}
\put(00,20){\vector(1,0){35}}

\put(70,20){\circle*{3}}
\put(70,20){\line(-1,0){70}}
\put(70,00){\circle*{3}}
\put(70,00){\vector(-1,0){35}}

\put(110,00){\circle*{3}}
\put(110,20){\line(1,0){20}}
\put(130,10){\oval(20,20)[r]}
\put(130,00){\line(-1,0){20}}
\put(110,20){\circle*{3}}
\put(110,20){\vector(1,0){15}}
\put(160,20){\vector(1,0){10}}

\put(180,20){\circle*{3}}
\put(180,20){\line(-1,0){20}}
\put(160,10){\oval(20,20)[l]}
\put(160,00){\line(1,0){20}}
\put(180,00){\circle*{3}}
\put(180,00){\vector(-1,0){15}}
\put(130,00){\vector(-1,0){10}}

\put(83,8){$\boldsymbol\Longrightarrow$}
\end{picture}
\ee
Longitudinal tension of the string forces the connecting part -- we refer to it as the ``zipper" -- to move longitudinally.
If it is made of a semicircular string piece with diameter $d$, then its acceleration is
\be a_{\parallel} = {4 \over \pi d}\,, \ee
and the relativistic motion with such acceleration in terms of rapidity and proper time is
simply given by
\be y_{zipper}=a_{\parallel} \tau\,. \ee
Since $\tau<\tau_{breaking}$ and $d\sim 1\,\fm \sim \tau_{breaking}$, one finds that a zipper can only move
by about 1 unit of rapidity during the time considered, of the total rapidity interval $2Y\sim 10$.
We thus conclude that there is no enough time to ``unzip" the string system.

\begin{figure}[t]
  \begin{center}
  \includegraphics[width=6cm]{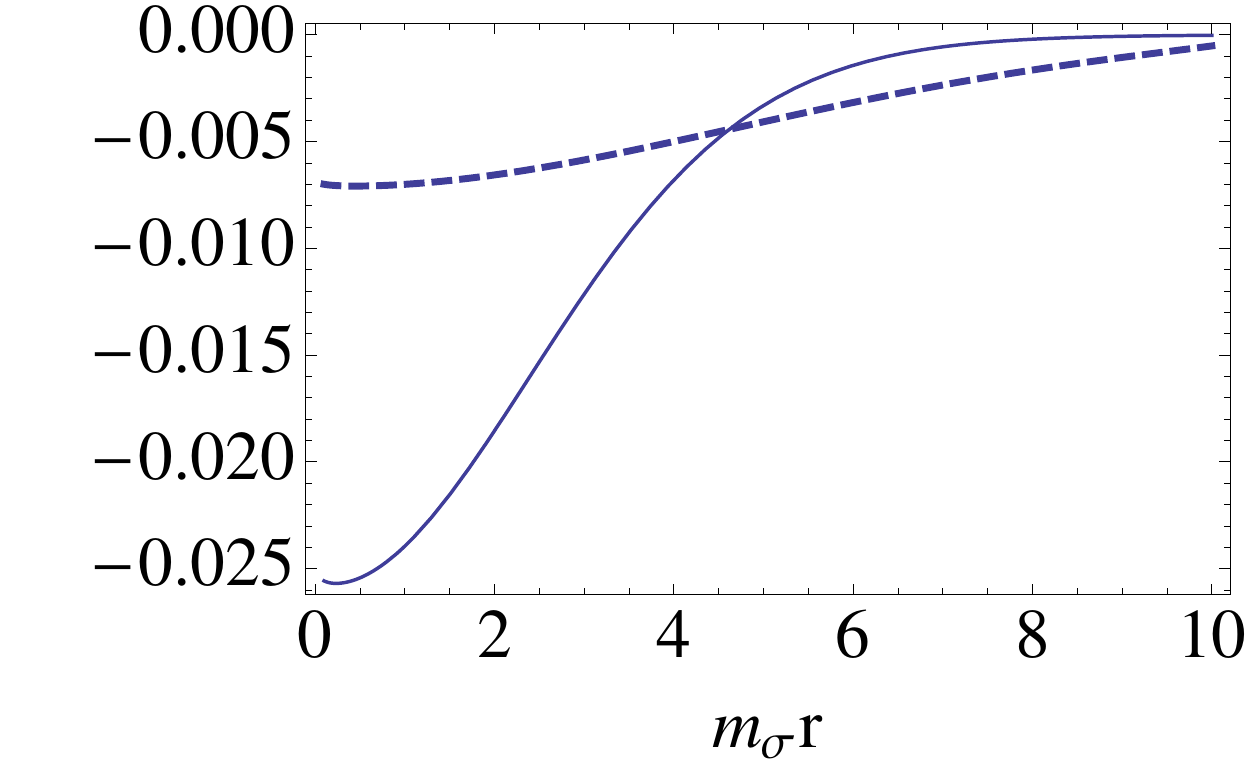}\\
  \caption{ (Color online) The mean field (normalized as explained in the text) versus the transverse radius in units of inverse $m_\sigma$. Dashed and solid curves correspond to the source radii $R=1.5$ and $0.7\, \fm$, respectively.}
  \label{fig_mean}
  \end{center}
\end{figure}

 \subsection{Mean field}

Assuming cylindrical symmetry, one can get the shape of the mean sigma distribution by
 solving the radial equation on the sigma field. We write it as
 \begin{align}
 \sigma''(r_\perp) + {1 \over r_\perp} \sigma'(r_\perp) -m_\sigma^2\sigma(r_\perp) =   \rho(r_\perp)\,,\label{sigma_eq}
 \end{align}
where $\rho(r_\perp)$ is the matter distribution in the transverse space.
Note that we have not included the coupling constant on the right-hand side or any normalization factors: this can be simply incorporated
into the solution once it is known, since Eq.~(\ref{sigma_eq}) is linear.
 We  use, for example, a Gaussian source, $\rho=\exp[-r_\perp^2/(2 R^2)]$.

At large distances the right-hand side of (\ref{sigma_eq}) is negligibly small, and the solution has the form
\begin{align}
  \sigma(r_\perp)=C\cdot K_0(m_\sigma r_\perp)\,,
  \end{align}
which can be used to fix asymptotics of the numerical solution at large $r$.
If the integration is performed starting from a large $r$ downwards, then the generic solution blows up at small $r$, unless
the constant $C$ is specially tuned. In Fig.~\ref{fig_mean} we show two such  solutions,
with tuned constants $C=3757.21$ and $42.37$ and radii $R=1.5$ and $0.7\, \fm$, respectively
(the solutions are rescaled in the plot, so that the integral of the source is normalized to 1).
These two radii are supposed to exemplify the ``spaghetti" transverse size before and after a collapse:
as one can see from the figure, the depth of the sigma potential well increases roughly by a factor of 5 between these two cases.
 This is more than enough to completely cancel chiral symmetry breaking around the after-collapse system.

\section{Molecular dynamics study}
\subsection{Initialization for central $pA$ and peripheral $AA$}

 \begin{figure}[t]
  \includegraphics[width=6cm]{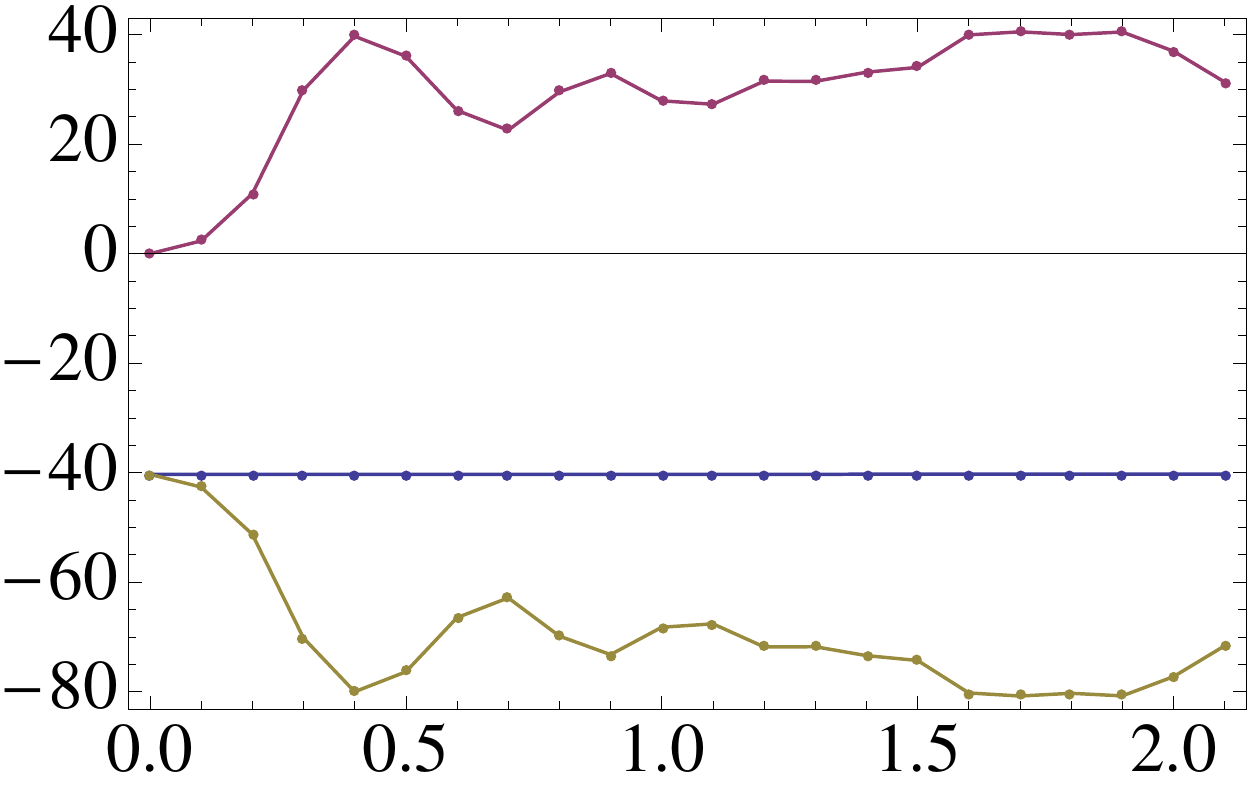}
  \caption{(Color online).  The (dimensionless) kinetic energy (upper curve) and potential energy (lower curve) of the system
  for the same example as shown in Fig.~\ref{fig_pA}, as a function of time $t\, (\fm/c)$. The horizontal line with points is their sum, $E_{tot}$, which is conserved.}
  \label{fig_energy}
\end{figure}
 \begin{figure}[]
   \includegraphics[width=6cm]{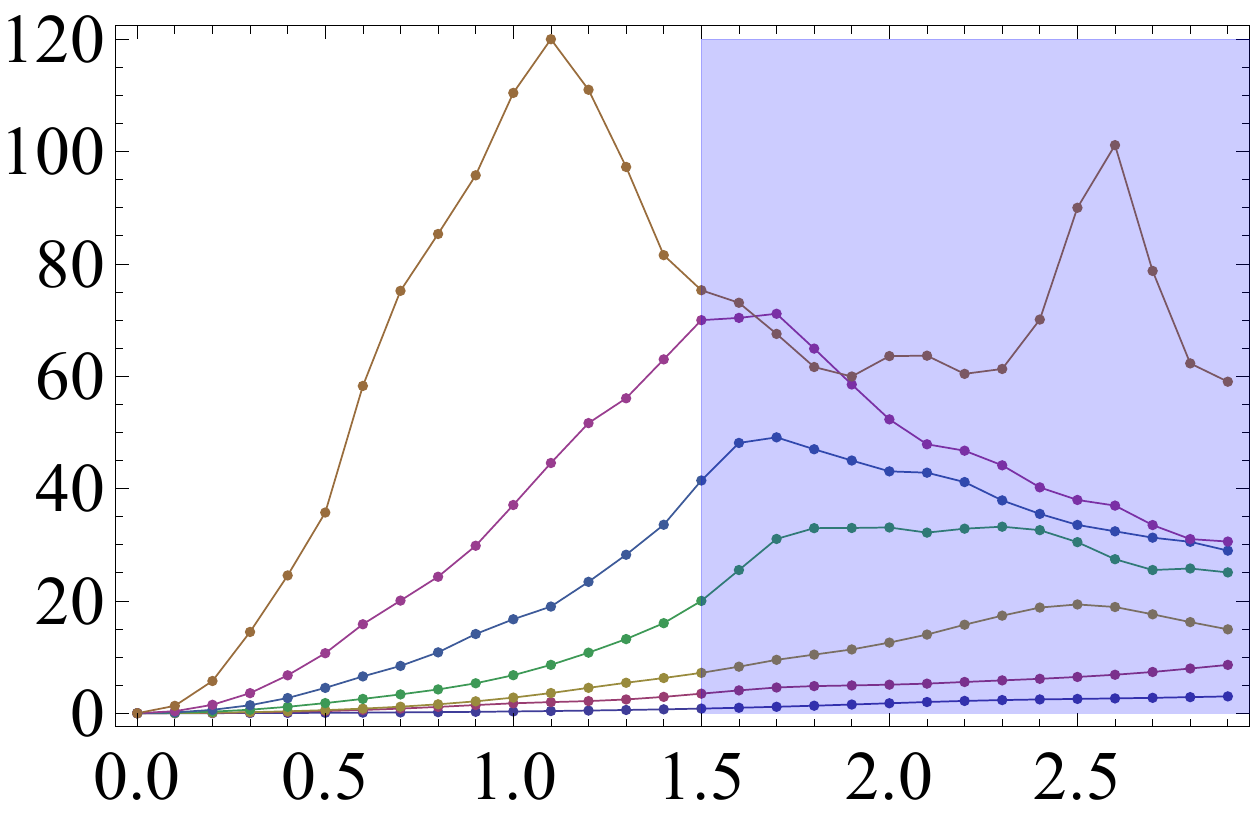}
   \caption{(Color online). Kinetic energy (dimensionless) versus the simulation time ($\fm/c$), for few $pA$ runs with $N_s = 50$.
   The seven curves (bottom-to-top) correspond to increasing coupling constants $g_N \sigma_T=0.01, 0.02, 0.03, 0.05, 0.08, 0.10$, and $0.20$. The shaded region at the right corresponds to the time which is considered to be too
   late for strings to exist, due to their breaking.}
   \label{fig_kinetic}
\end{figure}

  \begin{figure*}[t]
  \includegraphics[width=4cm]{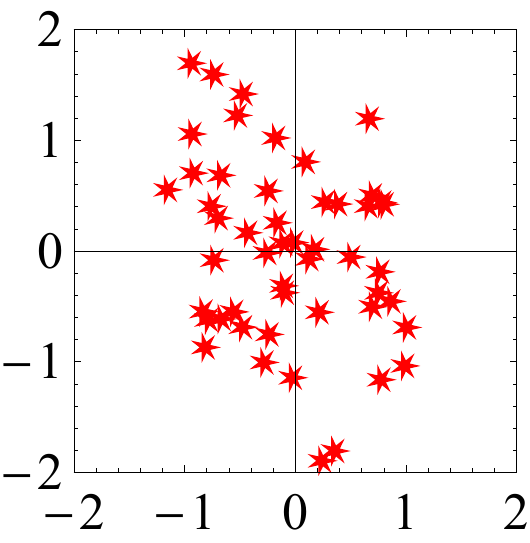}\hspace{0.2cm}
  \includegraphics[width=4cm]{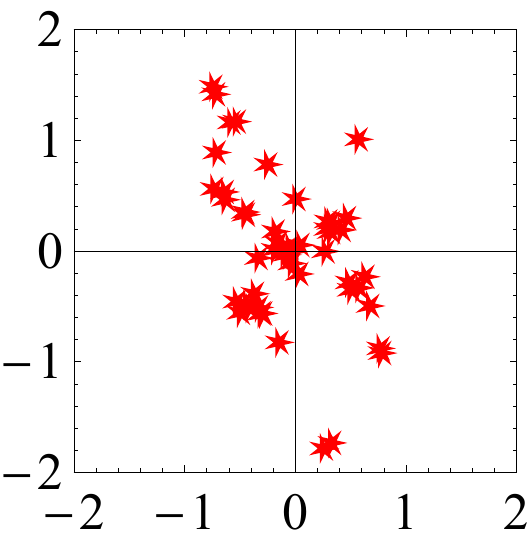}\hspace{0.2cm}
  \includegraphics[width=4cm]{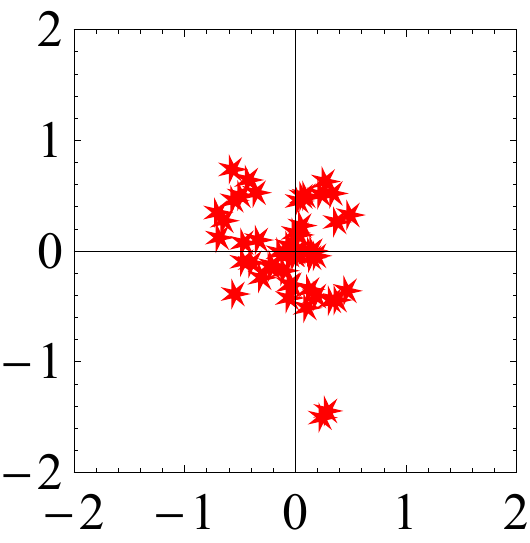}\hspace{0.2cm}
  \includegraphics[width=4cm]{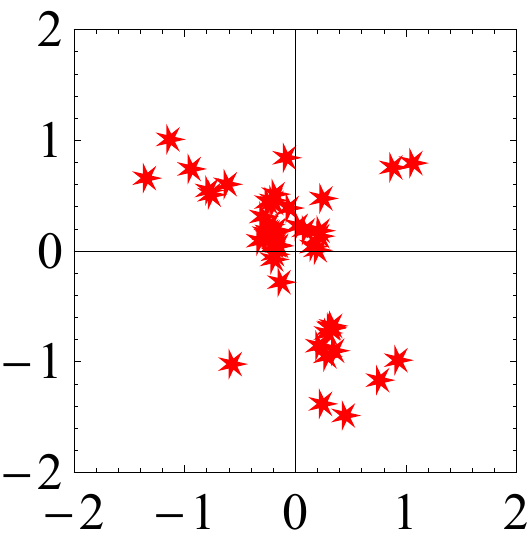}
  \caption{(Color online) Example of changing transverse positions of the 50 string set: the four plots correspond to
  one initial configuration evolved to times $\tau=0.1,\,0.5,\,1$, and $1.5\, \fm/c$. Distances are given in femtometers, and $g_N \sigma_T=0.2$.}
  \label{fig_pA}
\end{figure*}

   \begin{figure*}[t]
  \includegraphics[width=3cm]{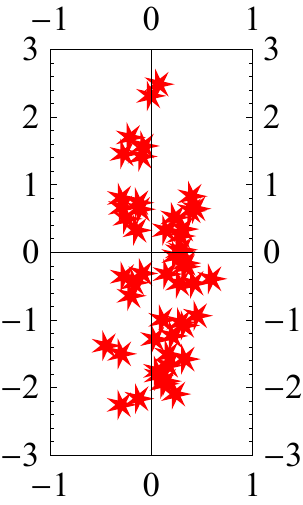}\hspace{0.9cm}
  \includegraphics[width=3cm]{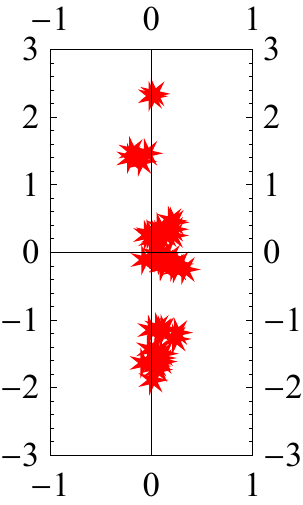}\hspace{0.9cm}
  \includegraphics[width=3cm]{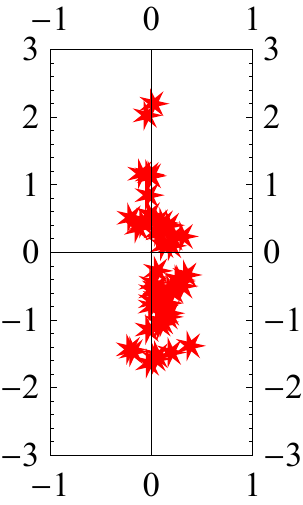}\hspace{0.9cm}
  \includegraphics[width=3cm]{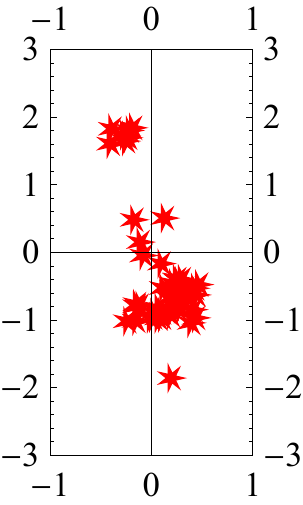}
  \caption{(Color online) Example of peripheral $AA$ collisions, with $b=11\, \fm$, $g_N \sigma_T=0.2$, and the 50-string set.
The four plots of the string transverse positions $x,y$ (fm) correspond to
  times $\tau=0.1,\,0.5,\,1.$, and $2.6\,\fm/c$.}
  \label{fig_AA}
\end{figure*}

To simulate central $pA$ we first select the number of participant (or ``wounded") nucleons
$N_p=5,10,15,20$, and $25$ and select their random positions in the transverse plane.
The numbers correspond to $p$ moving along the diameter of $Pb$ as discussed above, while variation in
the number roughly corresponds to expected fluctuations.

 The azimuthal angle is random, with the uniform distribution. The
distance from the origin, $b$ (i.e., the center of the original proton), is taken with the
weight given by the elastic scattering amplitude profile:
\begin{align}
\dd P=F(b)\, b\, \dd b\,.
\end{align}
We used the profile function from the Bourrely-Soffer-Wu (BSW) model~\cite{Bourrely:2012hp} [see their expressions (13-15)], which fitted well to the LHC data.
Because the profile function is flat, $F(b)\approx 1$ for $b<0.4\, \fm$, the density of strings is approximately constant at the center.
The corresponding initial energy density in this region,
\begin{align}
\epsilon_0=\sigma_T n_s \,,
\end{align}
 ranges from about $\epsilon_0=2$ to $9 \, \GeV/\fm^3$. The lower value, if equilibrated, would belong to the mixed phase.
The upper value
 is inside the QGP domain, but barely.  So, even if equilibration of this state happened, one would  not expect a hydro explosion, as its equation of state is quite soft.
 As we show below, collective string effects could increase these numbers by roughly an order of magnitude
 and, thus, change the conclusion.

Peripheral $AA$ are modeled in the standard Glauber way, except that we take the number of participants
being in exactly the same bins, namely, $N_p=5,10,15,20$, and $25$, for comparison.
 \subsection{Peripheral AA\label{sec_peripheral_AA}}

 Let us briefly describe how we choose the initial string configurations in this case.
 The relation between the impact parameter of two nuclei $b_{AA}$ and the number of participant nucleons $N_p$
 is given by the standard Glauber simulation, with $\sigma_{tot}=120 \, \mb$. We
 then select bins with a certain number of $N_p$, and pick out several random events from an ensemble.
 These are evolved using the molecular dynamics code.

    The Pomeron undergoes a tunneling (Euclidean) stage, from which it appears as two strings (with opposite
    color fluxes). At time t=0 the strings are transverse to the beam, with their length given by the impact parameters between
    nucleons $b_{NN}$. The strings are also separated in the transverse plane by
    \begin{align}
    \delta x_\perp = {\pi \over T_{eff} }\,,
    \end{align}
 where $T_{eff}$ is the effective temperature introduced in  \cite{Shuryak:2013sra}. While, in principle,
 it depends on $b_{NN}$ and the
 logarithm of the collision energy, we did not include these dependencies and use a randomly oriented vector of length
 $0.2\,\fm$.

 Since the end points of each string are at large rapidities $\sim \pm Y$, at $t>0$ strings interpolate linearly between these
 points. Since most measurements are done at midrapidity, the transverse location of the string is at the midpoint
 between the two nucleons. Thus we do not need to model $b_{NN}$ and its distribution.

 \begin{figure*}[!ht]
  \centering
  \includegraphics[width=7cm]{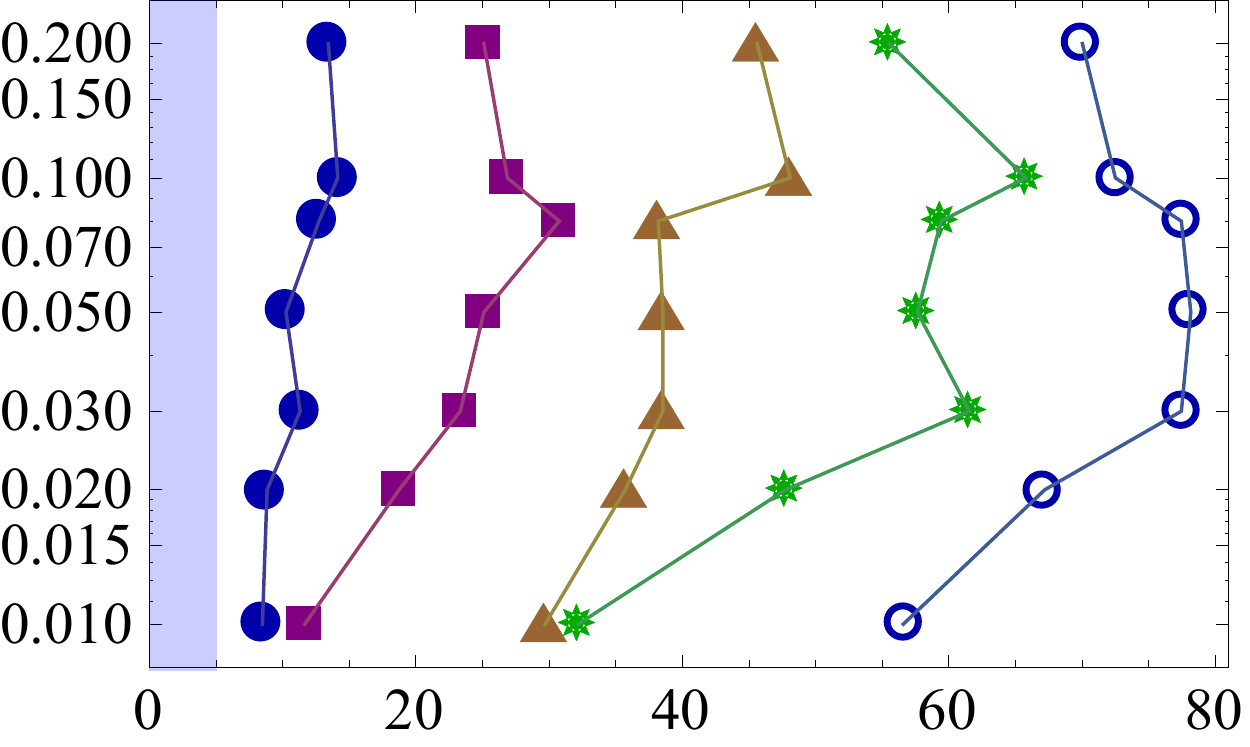}\hspace{1.2cm}
  \includegraphics[width=7cm]{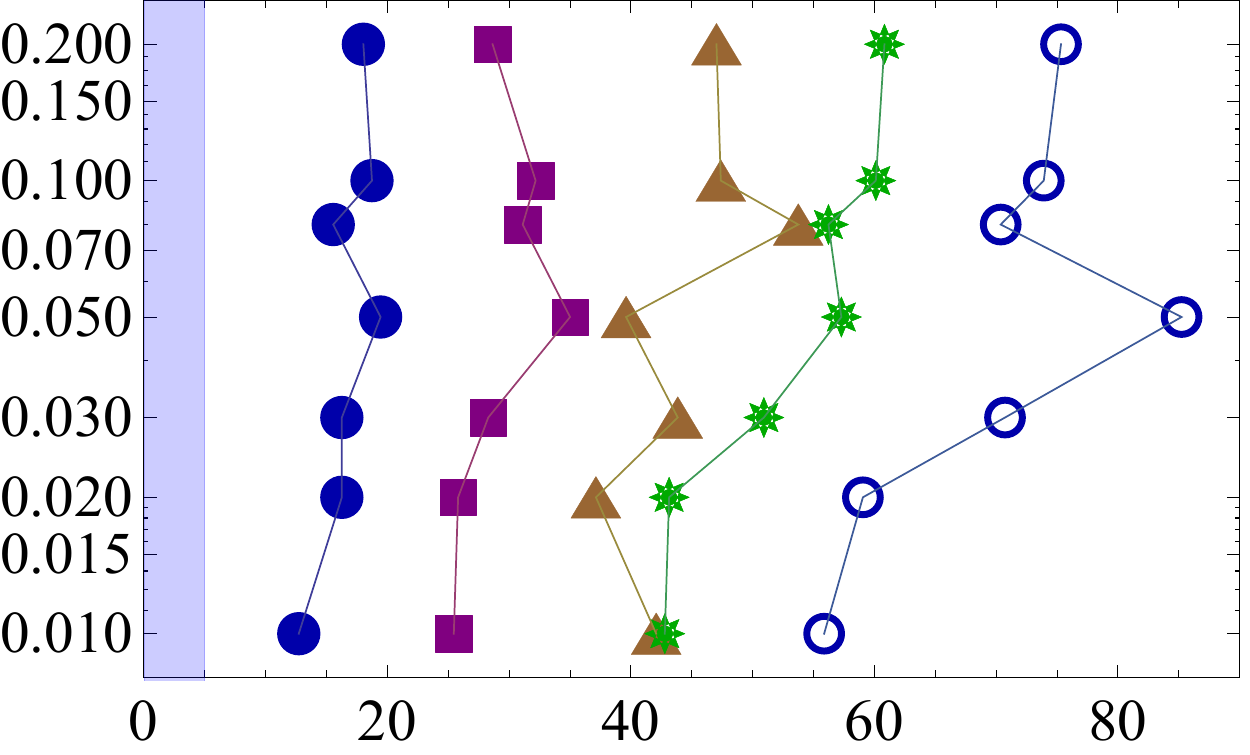}
  \caption{(Color online)  The left plot is for central $pA$, the right one -- for peripheral $AA$ collisions.
  The vertical axis is the effective coupling constant $g_N \sigma_T$ (dimensionless). The horizontal
  axis is the maximal energy density $\epsilon_{max}$ ($\GeV/\fm^3$) defined by the procedure explained in the text.
  Five sets shown by different symbols correspond to string number $N_s=10,20,30, 40,50$, left to right respectively.}
  \label{fig_ppp}
\end{figure*}

 \subsection{Time evolution}

As discussed above, the strings can be viewed as a 2D gas of particles (in the transverse plane) with unit masses at positions $\vec{r}_{i}$. The
forces between them are given by the derivative of the energy, (\ref{eqn_pot}), and so
\be \ddot{\vec{r}}_i= \vec{f}_{ij}= {\vec{r}_{ij}\over \tilde{r}_{ij}} (g_N \sigma_T) m_\sigma  2 K_1(m_\sigma \tilde{r}_{ij}) \ee
with $\vec{r}_{ij}=\vec{r}_{j}-\vec{r}_{i}$ and ``regularized" $\tilde{r}$ (\ref{eqn_tilde_r}).

In our simulations we used a classical molecular dynamics code based on the double-precision CERNLIB solver DDEQMR.
In Figs.~\ref{fig_pA} and \ref{fig_AA} we show an example of one particular configuration with $N_s=40$. In order to study a longer time evolution, we took a somewhat larger coupling, $g_N\sigma_T = 0.2$. As shown in Fig.~\ref{fig_energy}
   the conservation of the (dimensionless) total energy
\be E_{tot} =\sum_i  {v_i^2 \over 2}- 2 g_N \sigma_T \sum_{i>j} K_0(m_\sigma r_{ij})  \ee
is indeed observed: its accuracy is about $10^{-4}$. An even higher accuracy is observed for the total momentum
(which remains 0).

The evolution consists of two qualitatively distinct parts: (i) early implosion, which converts the potential energy into the kinetic one, which has its peak when a
fraction of the particles ``gravitationally collapses" into a tight cluster; and (ii) subsequent approach to
a ``mini-galaxy" in virial quasi-equilibrium.
To illustrate better the first stage of the motion we took a number of screenshots: a typical case is shown in
   Fig.~\ref{fig_pA}.  Starting from various initial configurations we occasionally see more complicated scenarios realized, e.g.,
   two minigalaxies flying away from each other.

 One can see that the total kinetic energy shows a peak and then oscillates. Its average over time approaches  some mean value, which of course should be related to the ``virial'' value
    \be 2\langle E_{kin} \rangle = \left\langle \sum_i \vec{r}_i {\partial U \over \partial  \vec{r}_i} \right\rangle \ee
as time goes to infinity. (This is the standard outcome of molecular dynamics studies, e.g., stars in galaxies
have a similar quasi-equilibrium).

   Simulations for peripheral $AA$ show a particular feature. As exemplified in Fig.~\ref{fig_AA}, the initial strong deformation
   of the system -- its $y$-direction size is much larger than that in the $x$ direction -- the collapse occurs in two stages.
   First, one observes a rapid 1D collapse along the $x$ axes, followed by a much slower collapse
   along the $y$ direction. If the simulation runs long enough, the resulting cluster  becomes, of course, isotropic.

\begin{figure}[b]
  \begin{center}
  \includegraphics[width=6cm, valign=t]{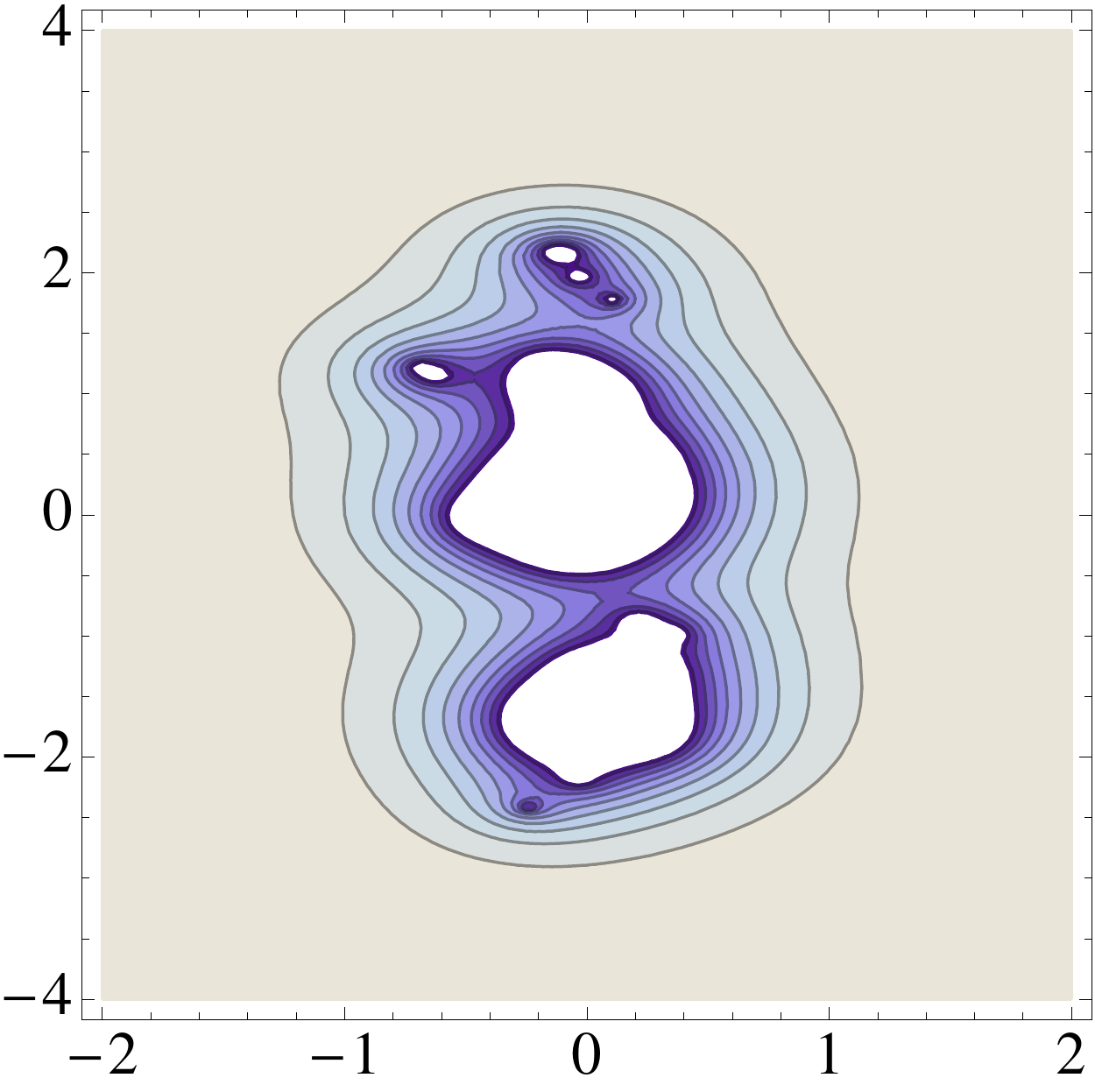}\hspace{0.5cm}\includegraphics[width=1.05cm, valign=t]{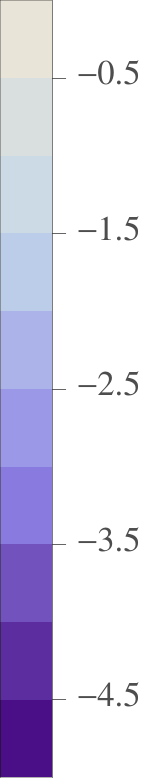}
  \caption{ (Color online)  Instantaneous collective potential (in units of $2 g_N \sigma_T$) for an $AA$ configuration with $b=11\,\fm$, $g_N \sigma_T = 0.2$, $N_s=50$ at the moment in time $\tau = 1\,\fm/c$. White regions correspond to the chirally restored phase.}
  \label{allAA_pot}
  \end{center}
\end{figure}

\subsection{Results}
We simulated similar time evolutions for ensembles  of randomly generated initial conditions.
Systematic results were organized as follows. We have ensembles of 10 independent runs for each set of parameters, string numbers $N_s=10,\, 20,\, 30,\, 40$, and $50$, coupling constants $g_N \sigma_T=0.01,\, 0.02,\, 0.03,\, 0.05,\, 0.08,\, 0.1$, and $0.2$, and two different initializations, corresponding to the central $pA$ and peripheral $AA$ cases.

The dependence of the evolution timescale on the value of the coupling is shown in  Fig.~\ref{fig_kinetic}.
While the value itself grows substantially with the coupling, the ``implosion time" (the location of the first peak)
depends on the coupling more gradually.

Of the many possible observables we selected the local density in the generated clusters $\epsilon_{max}$  defined by the following procedure.
As the first step, we find the location of {\em the most rapidly moving particle}, resembling early searches for the location of the black hole at our galaxy center.
After it is found, its position is taken as the cluster center, and the number of particles inside the circle of fixed radius $r_0= 0.3 \, \fm$
is used to calculate the maximal 2D density $n_{max}$.
   The results are converted to the maximal energy density of a run by
   \begin{align}
   \epsilon_{max}=\sigma_T n_{max}
   \end{align}
   and averaged over the runs.

    The output is shown in Fig.~\ref{fig_ppp} as the maximal energy density reached (during the proper time $\tau<2 \,\fm/c$).
  The main result is that the implosion of the system
  produces values which  are significantly higher than those at the initial time $\tau=0$, namely
  $\epsilon_0=2$ to $9 \, \GeV/\fm^3$ for these sets.

  While the rate of evolution depends on the strength of the coupling, the maximal energy density reached
is much less sensitive to it.
As one can see from this, for a small number of strings, $\sim 10$, there is no dependence on the coupling in the range selected:
these are too small to create any effect. However, as $N_s>30 $, the  coupling becomes important: it increases the density
by a significant factor, reaching values as large as $ \epsilon_{max}\sim 80\, \GeV/\fm^3$.

As such a high energy density is being reached, the string description of the system can no longer be maintained.
The kinetic energy is transferred into multiple
string states, and the strings become highly excited. If the system fully equilibrated into the sQGP, the temperature would be about $T_i\sim 500 \MeV \sim 3\,T_c$, enough to generate a very robust hydro explosion.

Finally, in Fig.~\ref{allAA_pot} we show an example of the instantaneous collective potential produced by the strings in the transverse plane. The white regions correspond to potential values smaller than $-5\cdot 2 g_N \sigma_T (\fm^{-1}) \approx -400\,\MeV$; i.e., chiral symmetry can be completely restored in these regions.
A large gradient of this potential at its edge can cause quark pair production, similar to the Schwinger process in an electric field: one particle may flow outward and
one fall into the well. This phenomenon is a QCD analog to Hawking radiation at the black hole horizon.
An analogous picture was also considered in \cite{Kharzeev:2005iz} and \cite{Castorina:2007eb}.
The final ellipticity of the induced elliptic flow will be studied elsewhere.

\section{Summary and outlook}
In this work we have discussed collective interactions between QCD strings in a ``spaghetti" configuration,
created in ``central" $pA$ and peripheral $AA$ collisions. We first provided an experimental overview, concluding that  at least three observables -- multiplicity distribution, radial flow and elliptic flow -- show the onset of a different regime at the string number $N_s \sim 30$.
Although this number may appear to be large, one can see that, naively, the produced system remains sufficiently dilute.
In particular, under this condition the chiral condensate is expected to be modified only at the level of 10\% or so.

Next we formulated a model of the string-string interaction induced by the $\sigma$ meson exchange and matched it to the lattice data. We performed a molecular dynamics simulation of  the string motion  in the 2D  transverse plane.
We observed collective implosion of the ``spaghetti" configurations and listed parameters of the string interaction
which may cause the transition. The range of the string numbers is chosen to correspond to the transition in experiments.

 One may argue that the string description must break down, as the string density (and hence the energy density of the system) is increased by a significant factor due to the implosion. It is expected that
it undergoes a rapid equilibration into the QGP phase, which then explodes hydrodynamically, in agreement with previous studies.
We argue that the proposed ``spaghetti implosion" is the crucial  piece of the puzzle, explaining  the change in the dynamics.

We mentioned in Sec.~I that, in the AdS/CFT  vocabulary,
 thermal fireballs of deconfined matter are dual to certain 5D black holes and that the
attractively interacting and collapsing system of QCD strings we discuss must be a QCD
analog to AdS/CFT black hole formation.
As an outlook we would like to mention further developments of this correspondence, in the holographic AdS/QCD
framework. 

In AdS/QCD models the string-string interaction is also attractive, mediated by massless dilaton and graviton.
There is no need for additional parameters, like our $\sigma$-string coupling, as that is already defined
by the model action. It would be interesting to investigate under what conditions multistring
 implosion should happen, and whether this indeed leads to gravitational horizons and black holes,
 not just a higher density of strings. If so,
the holographic framework uniquely connects the black hole production to many famous
effects.
The collective sigma field displayed in Fig.~\ref{fig_mean} should presumably lead to the
particle pair production at its edge -- the QCD analog to Hawking radiation.
And the amount of entropy produced in multistring annihilation/excitation should be
somehow analogous to the universal Bekenstein entropy of a black hole.

{\bf Acknowledgements.}
We would like to thank  Dmitri Kharzeev, Frasher Loshaj and Ismail Zahed for useful discussions.
This work was supported in part by the U.S. Department of Energy under Contract No. DE-FG-88ER40388.

\end{document}